\documentclass[prb,aps,twocolumn,reprint,amsmath,amssymb,showpacs,superscriptaddress]{revtex4}

\usepackage{bm}
\usepackage{dcolumn}
\usepackage{graphicx}
\usepackage{amsfonts}
\usepackage{amsmath}
\usepackage{amssymb}
\usepackage{subfigure}
\usepackage{color}
\newcommand{\beq}{\begin{equation}}
\newcommand{\eeq}{\end{equation}}
\newcommand{\beqarray}{\begin{eqnarray}}
\newcommand{\eeqarray}{\end{eqnarray}}
\newcommand{\Ref}[1]{Ref.~\onlinecite{#1}} 
\newcommand{\eq}[1]{Eq.~(\ref{#1})} 
\newcolumntype{.}{D{.}{.}{-1}}
\begin{document}
\title{Theory for Magnetism and Triplet Superconductivity in LiFeAs}

\author{P. M. R. Brydon}
\email{brydon@theory.phy.tu-dresden.de}
\affiliation{Institut f\"ur Theoretische Physik, Technische Universit\"at Dresden, 01062 Dresden, Germany}
\author{Maria Daghofer}
\email{m.daghofer@ifw-dresden.de}
\affiliation{IFW Dresden, P.O. Box 270116, 01171 Dresden, Germany}
\author{Carsten Timm}
\affiliation{Institut f\"ur Theoretische Physik, Technische Universit\"at Dresden, 01062 Dresden, Germany}
\author{Jeroen van den Brink}
\affiliation{IFW Dresden, P.O. Box 270116, 01171 Dresden, Germany}

\date{\today}

\begin{abstract}
Superconducting pnictides are widely found to feature spin-singlet
pairing in the vicinity of an antiferromagnetic phase, for which nesting
between electron and hole Fermi surfaces is crucial. LiFeAs differs from
the other pnictides by (i) poor nesting properties and (ii) unusually
shallow hole pockets. Investigating magnetic and pairing instabilities
in an electronic model that incorporates these differences, we find
antiferromagnetic order to be absent. Instead we observe almost
  ferromagnetic fluctuations which drive an instability
toward spin-triplet $p$-wave superconductivity.
\end{abstract}

\maketitle

\section{Introduction} 

The family of pnictide superconductors has generated large interest 
during the last two years, because their superconductivity (SC) is generally
thought to be unconventional, i.e., not due to phonons, and because
some pnictides have high transition temperatures 
reminiscent of cuprates. The details of the pairing are far from being
clarified, and may vary significantly between the different families; 
for two recent reviews see Refs.~\onlinecite{Paglione:2010p2493,pnict_rev_johnston}.
A plausible scenario for the pairing
mechanism involves exchange of virtual spin fluctuations, as SC is 
often observed in the 
vicinity of an antiferromagnetic (AF) spin-density-wave (SDW) state. In the
  1111 and 122 families, the SDW is apparently driven by excellent nesting of electron
  and hole Fermi pockets. 
 When doping reduces the nesting, SDW order is weakened, but AF
fluctuations remain and are believed to provide the pairing for the
Cooper pairs.~\cite{Mazin:2008p1695,chubukov:134512}
 Together with the high transition temperatures
  observed in some 
  pnictides, such a scenario raises hopes that pnictides might shed light onto
similarities -- or differences -- between spin-fluctuation superconductors like
heavy-fermion systems, with typically 
lower transition
temperatures,~\cite{Mathur:1998p2498,Jourdan:1999p2499} and the more
enigmatic cuprates.

The recently-discovered pristine superconductor LiFeAs differs in key
  respects from the most commonly studied pnictide families. Doping is not
necessary for
superconductivity;~\cite{Tapp:2008p2485,LiFeAS_no_mag_2009,Borisenko:2010p2488} 
the nesting between hole and electron pockets is rather poor and consequently
an SDW is not observed. 
Moreover, compared to other pnictide families, LiFeAs has much shallower
  hole pockets around the center of the Brillouin
zone.~\cite{Borisenko:2010p2488,eschrig_tb,LeFeAs_surf} It was pointed
  out that the flat top
of these pockets implies a large density of
states,~\cite{Borisenko:2010p2488} which one can  expect to promote
ferromagnetic (FM) fluctuations. Indeed, recent nuclear magnetic
resonance (NMR) measurements indicate a relatively high susceptibility
within the $a$-$b$ planes. Remarkably, the Knight shift does not
change at the superconducting transition for magnetic fields
  perpendicular to the $c$ axis,~\cite{LiFeAs_triplet} which is evidence
    for triplet
superconductivity. This discovery relates a member of the pnictide family to
yet another family with unconventional SC, the ruthenates.~\cite{Mackenzie:2003p2495}

 In this paper we 
model 
the distinguishing electronic features of LiFeAs, i.e., the absence of
nesting and the presence of shallow hole pockets. 
We examine the magnetic properties of this Hamiltonian by combining a mean-field
approximation and the  random-phase approximation (RPA). We find that the
magnetic fluctuations are dominated by almost FM processes
across the smaller hole pocket.
We proceed to investigate what types of Cooper
pairs can be stabilized by the spin and charge fluctuations. We use the
same approach that has been used to treat other pnictides, and which
yields  AF order and singlet pairing in the 1111 and 122
families.~\cite{graser_5b,Kemper:2010p2492,Graser:2010p2497,ql_luo_2010} In the
case of LiFeAs, however, we find that the almost FM fluctuations can drive
triplet $p$-wave superconductivity. Throughout, we discuss the connection of
our work to experimental results.

The possibility of triplet pairing was already discussed in the early days of pnictide research. 
The degeneracy of the $xz$ and $yz$ orbitals that dominate the Fermi
surface (FS) permits spin-triplet and orbital-singlet pairs with
(nodal)  $s$- or $d$-wave-like
character.~\cite{zhou,Wan:2009p1056,wang,FCZhang,Daghofer:2008p1970,Xu:2008p2490} 
Triplets with $p$-wave
symmetry, 
stabilized by large Hund's rule coupling, were reported for a three-band model and involved mostly the
electron pockets.~\cite{plee} 
The mechanism proposed in this paper, however, is very different and
  specific to LiFeAs, originating from the high
density of states due to the rather flat `roofs'
of the shallow \emph{hole} pockets.

\begin{figure}
\begin{minipage}{0.65\columnwidth}
\subfigure{\includegraphics[width=\textwidth]{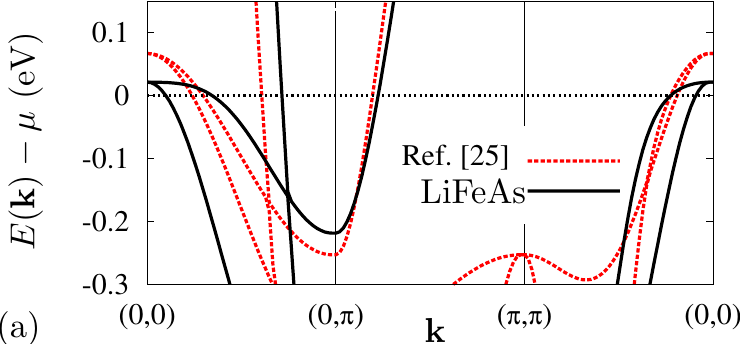}\label{fig:bands}}
\end{minipage}
\begin{minipage}{0.33\columnwidth}
\subfigure{\includegraphics[width=\textwidth]{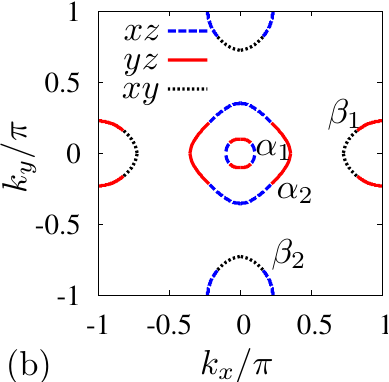}\label{fig:fs}}
\end{minipage}
\caption{(Color online)  (a) Band structure of (dashed lines) the
  three-band model  Eq.~(\ref{eq:H0k}) from Ref.~\onlinecite{Daghofer_3orb} and (solid lines) the
  model for LiFeAs. (b) Fermi surface and dominant orbitals for the
  LiFeAs model; $\alpha_1$ and $\alpha_2$ are hole pockets, 
  $\beta_1$ and $\beta_2$ are electron pockets. Parameters:   
  $t_1 =0.02$, $t_2 =0.12$, $t_3 =0.02$, $t_4=-0.046$, $t_5=0.2$, $t_6
  =0.3$, $t_7 = -0.15$, $t_8 = -t_7/2$, $t_9 = -0.06$, $t_{10}=
  -0.03$, $t_{11} =  0.014$, $\Delta_{xy} = 1$, $\mu=0.338$. The
  filling is four electrons per site, energies are in eV. 
\label{fig:bands_fs}}
\end{figure}

\section{Model}

In  order to capture the most relevant states around the FS, we model the two concentric hole pockets and the two
electron pockets, which can be achieved using three orbitals per iron
ion. We use a one-iron unit cell, with ${\bf k}$
running over $-\pi<k_x,k_y\le\pi$ in the corresponding extended
Brillouin zone, and the hole pockets are then around $\Gamma = (0,0)$,
while the electron pockets are found at $X=(\pi,0)$ and $Y=(0,\pi)$.
The three orbitals with the largest weight in these three pockets
are the $xz$, $yz$, and $xy$ orbitals, denoted
by indices 1, 2 and 3.
The  momentum-dependent non-interacting Hamiltonian is given by
\begin{eqnarray}\label{eq:H0k}
H_0(\mathbf{ k}) &=& \sum_{\mathbf{ k},\sigma,\mu,\nu}
T^{\mu,\nu} 
(\mathbf{ k}) 
d^\dagger_{\mathbf{ k},\mu,\sigma} d^{\phantom{\dagger}}_{\mathbf{ k},\nu,\sigma}\;,
\end{eqnarray}
where $d^\dagger_{\mathbf{ k},\mu,\sigma}$
($d^{\phantom{\dagger}}_{\mathbf{ k},\mu,\sigma}$) creates
(annihilates) an electron in orbital $\mu$ with momentum ${\bf k}$ and
spin $\sigma$. The elements of the hopping matrix $T^{\mu,\nu} 
(\mathbf{ k}) $ are 
\begin{align}
T^{11/22} &= 2t_{2/1}\cos  k_x +2t_{1/2}\cos  k_y +4t_3 \cos  k_x 
\cos  k_y \nonumber \\
 &\quad \pm 2t_{11}(\cos 2k_x-\cos 2k_y)- \mu,  \\
T^{33} &= \Delta_{xy} + 2t_5(\cos  k_x+\cos  k_y) +4t_6\cos  k_x\cos k_y \nonumber\\
       &\quad  + 2t_9(\cos 2k_x+\cos 2k_y) \nonumber  \\
       &\quad + 4t_{10}(\cos 2k_x \cos k_y + \cos k_x \cos 2k_y)- \mu,  
 \\
T^{12} &= T^{21} =-4t_4\sin  k_x \sin  k_y,  \\
T^{13} &= \bar{T}^{31} = 2it_7\sin  k_x + 4it_8\sin  k_x \cos  k_y,  \\
T^{23} &= \bar{T}^{32} =  2it_7\sin  k_y + 4it_8\sin  k_y \cos  k_x,  
\end{align}
where a bar on top of a matrix element denotes the complex
conjugate. The parameters are given in the caption of Fig.~\ref{fig:bands_fs}.

\begin{figure}
\includegraphics[clip,width=0.9\columnwidth]{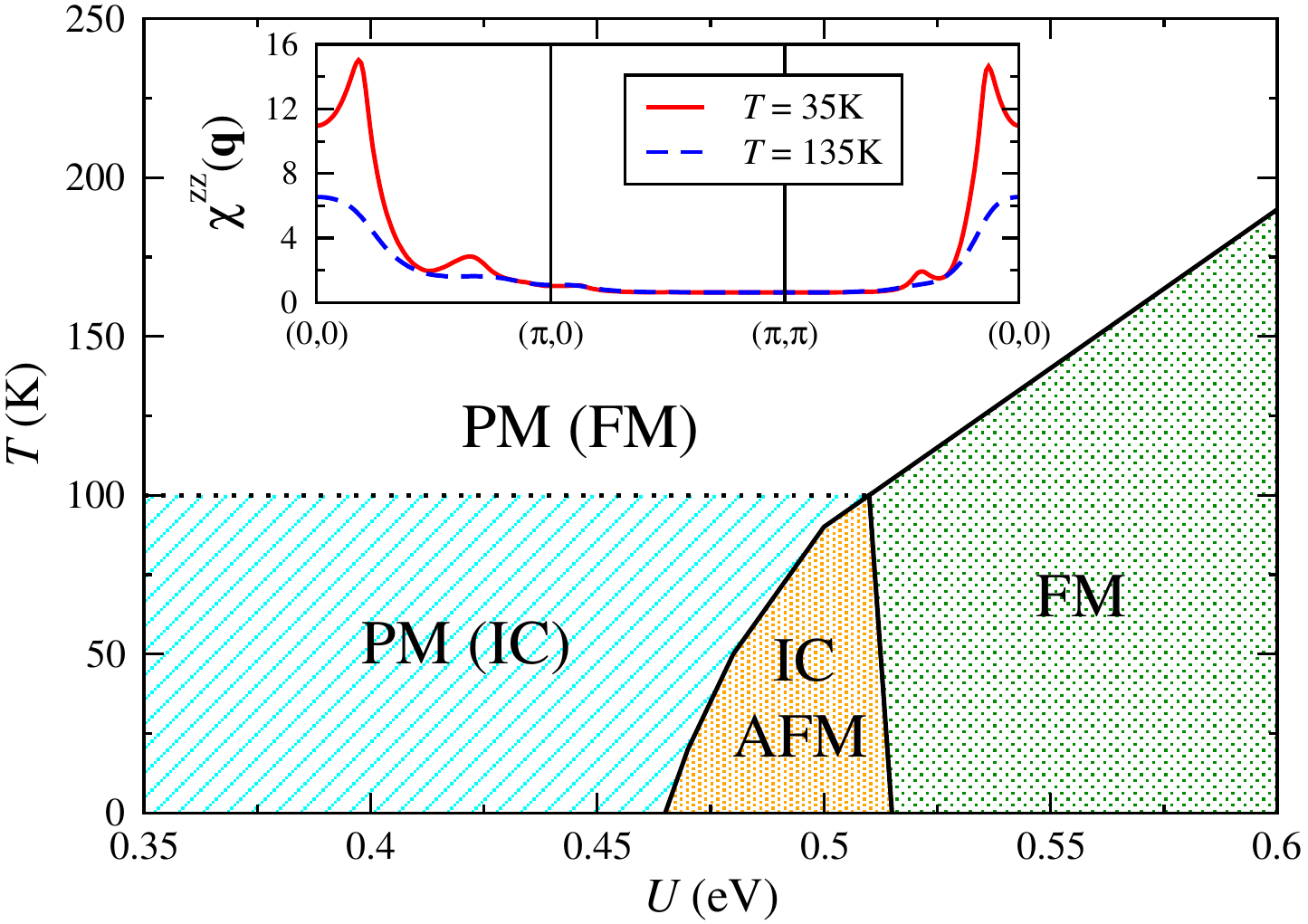}
\caption{(Color online) Magnetic phase diagram of the three-band model for
  LiFeAs. Inset: Magnetic susceptibility  $\chi^{zz}({\bf
    q})$ for $U=0.42$eV and
  $T=35\,$K and   $T=135\,$K.
 \label{fig:phases}}
\end{figure}

The one-particle bands defined by this kinetic energy are shown in
Fig.~\ref{fig:bands} for two parameterizations: one is the three-band model of
Ref.~\onlinecite{Daghofer_3orb}, where only hopping to first and
second neighbors is included. While this model already has rather shallow hole pockets
compared to {\it ab initio} bands for most pnictide compounds, it can reproduce spectral features of the AF
phase observed in many undoped
pnictides~\cite{Daghofer_3orb,orb_pol_FS} and has an instability to
singlet pairing.~\cite{ql_luo_2010} The second more refined model makes
use of longer-range hoppings in order to describe the bands of LiFeAs more
closely. The two hole pockets, which have mostly $xz$ and $yz$
character and are degenerate at $\Gamma$, are far shallower
than the electron pockets and have very different radii in accordance with
ARPES findings and {\it ab initio} density functional
calculations.~\cite{Borisenko:2010p2488,eschrig_tb,LeFeAs_surf} Both models
reflect the poor nesting of the electron and hole pockets in LiFeAs, see Fig.~\ref{fig:fs}. 
As will be discussed later, the
flat top of the hole pockets has a crucial impact on the model
properties; 
angle-resolved photo-emission (ARPES) data~\cite{Borisenko:2010p2488}
might suggest even flatter hole pockets.

We include all Coulomb interactions between electrons in the same Fe
atom: 
\begin{align}
H_{I} &= U\sum_{{\bf i},\nu}n_{{\bf i},\nu,\uparrow}n_{{\bf
    i},\nu,\downarrow} + V\sum_{{\bf
    i},\nu\neq\mu}\sum_{\sigma,\sigma'}n_{{\bf i},\nu,\sigma}n_{{\bf
    i},\mu,\sigma'}  \label{eq:Hint} \\
& - J\sum_{{\bf i},\nu\neq\mu}{\bf S}_{{\bf i},\nu}\cdot{\bf S}_{{\bf
      i},\mu} 
+ J'\sum_{{\bf i},\nu\neq\mu}d^{\dagger}_{{\bf
    i},\nu,\uparrow}d^{\dagger}_{{\bf i},\nu,\downarrow}d^{}_{{\bf
    i},\mu,\downarrow}d^{}_{{\bf i},\mu,\uparrow} \, .\nonumber
\end{align}
Here $n_{{\bf i},\nu,\sigma}$  (${\bf S}_{{\bf i},\nu}$) is the number
(spin) operator for the orbital $\nu$ on site ${\bf i}$. Invariance under rotation of the 
orbital degrees of freedom is ensured by setting $J = J'$ and $V = (2U -
5J)/4$. We only present results for $J=0.25U$, but we have
verified that other choices lead to similar results.

\section{Magnetism}

Following Ref.~\onlinecite{brydon_rpa_pnict_2010}, we construct the magnetic
  phase diagram by first using a mean-field ansatz restricted to FM and
  two-site AF phases~\cite{endn1}. Unlike  models of the
  1111 and 122 pnictides, here only FM and PM phases are stable.~\cite{brydon_rpa_pnict_2010}
The phase diagram is then refined by evaluating the static
transverse spin susceptibility $\chi^{-+}({\bf q})$ within the
RPA. This yields two important results: 
firstly, negative values of
$\chi^{-+}({\bf q})$ signal an instability toward a magnetically
ordered state (FM for ${\bf q}=0$, AF otherwise); secondly, 
in the PM phase the maximum of $\chi^{-+}({\bf q})$ indicates the
wavevector of the dominant spin fluctuations. 
In the PM phase
$\chi^{-+}({\bf q}) = 2\chi^{zz}({\bf q})$ is obtained by  
solving the Dyson equation $\hat{\chi}^{S} = \hat{\chi}^{(0)} +   
\hat{\chi}^{(0)}\hat{U}^{S}\hat{\chi}^{S}$ for the irreducible spin
susceptibility $\hat{\chi}^S$~\cite{endn2}.  The non-zero matrix
elements of $\hat{U}^{S}$ can be found in Refs.~\onlinecite{graser_5b,Kemper:2010p2492,Graser:2010p2497}. The Lindhard
function $\hat{\chi}^{(0)}$ is defined
\begin{gather}
\chi^{(0)}_{\nu,\nu',\mu,\mu'}({\bf q},i\omega_n) 
 =  -\frac{1}{N}\sum_{\bf k}\sum_{j,j'}
\frac{n_{F}(E_{j,{\bf k}}) - n_{F}(E_{j',{\bf k}+{\bf q}})}{E_{j,{\bf
      k}} - E_{j',{\bf k}+{\bf q}} - i\omega_n}\nonumber\\
\times\ u^{}_{j,\nu'}({\bf k})u^{\ast}_{j,\mu}({\bf
  k})u^{}_{j',\mu'}({\bf k}+{\bf q})u^{\ast}_{j',\nu}({\bf k}+{\bf q}) \label{eq:chi0} 
\end{gather}
where $\nu$, $\nu'$, $\mu$, $\mu'$ refer to the orbital,
$n_{F}(E)$ is the Fermi function, and $E_{j,{\bf k}}$ are the eigenvalues
of $H_{0}({\bf k})$ [Eq.~(\ref{eq:H0k})]. The coefficients $u_{j,\nu}({\bf k})$
transform the  diagonalizing (band) annihilation operators $\gamma^{}_{j,{\bf k}}$
(corresponding to $E_{j,{\bf k}}$) into the orbital
basis, i.e. $d^{}_{\nu,{\bf k}} = \sum_{j}u_{j,\nu}({\bf k})\gamma^{}_{j,{\bf
    k}}$. The static transverse spin
susceptibility is then written $\chi^{-+}({\bf q}) =
\sum_{\nu,\mu}\chi^{S}_{\nu,\nu,\mu,\mu}({\bf
  q},\omega=0)$. 

Figure~\ref{fig:phases} shows the phase diagram 
as a function of $U$ and temperature $T$. At low $U$, the model
remains PM down to $T=0\,$K, but the dominant
fluctuations change from FM to incommensurate (IC) at $T\approx100\,$K. As
$U$ is increased, the IC fluctuations can drive a transition to an IC-AF  
state, while at higher $U$ a FM phase is realized. 
The wave vectors characterizing the IC fluctuations in the low-$T$ PM
  state lie on a ring-shaped feature centered at ${\bf q}=(0,0)$, as can be
  seen
  in the example plotted as the inset in Fig.~\ref{fig:phases}.
The radius of the ring is exactly twice the radius of the
inner hole pocket, revealing that it originates from scattering diagonally
across this FS~\cite{endn3}. This scattering is strongly favored because of the identical
  orbital composition of diagonally separated parts of the FS.

Experimentally, a relatively large 
susceptibility is observed parallel to the $ab$-plane, which slightly decreases
with $T$.~\cite{LiFeAs_triplet} 
While an increasing susceptibility is expected for dominant FM
  fluctuations, this observation is consistent with `almost FM' IC
fluctuations, where the weight observed at $(0,0)$ in $\chi^{zz}({\bf
  q})$  no longer grows with lowering $T$ below the transition
from FM to IC fluctuations. It may even slightly decrease as the
ring feature becomes more pronounced, and our model shows such
behavior at low temperature $T\lesssim 50\,\textrm{K}$. 
These results indicate that 
LiFeAs is relatively near to a FM instability, 
which can be triggered by reducing the size of the inner hole pocket,
e.g. by electron doping. 

\begin{figure}
\subfigure{\includegraphics[width=0.47\textwidth]{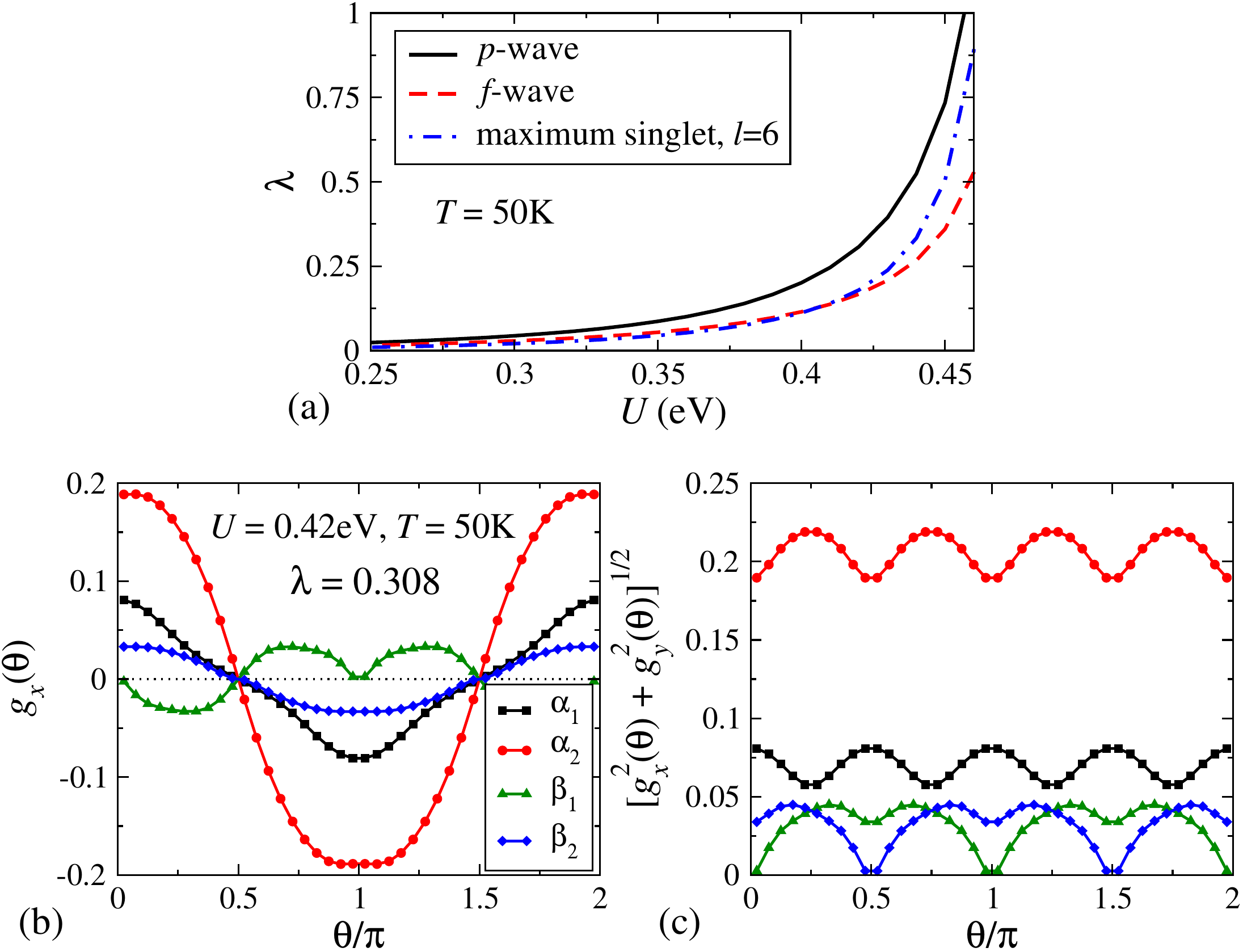}\label{fig:lambdas}}
\subfigure{\label{fig:gap_px}}
\subfigure{\label{fig:gap_comb}}
\caption{(Color online) Triplet pairing. (a)  Eigenvalues $\lambda$
    of the leading gap symmetries as a function of $U$.
  (b) Gap around the
  Fermi pockets for the $p_x$-wave pairing, which is a member of the
  degenerate pair that gives the dominant triplet $p$-channel. The
    winding angle about the FSs $\theta$ is measured with respect to the
    $k_x$-axis, taken in the anti-clockwise direction. (c) Gap magnitude of
  pairing states combining the degenerate $p_x$ and $p_y$ solutions
  of~\eq{eq:geigen}. $\alpha_1$, $\alpha_2$, $\beta_1$, and
    $\beta_2$ refer to the FS pockets, see Fig.~\ref{fig:fs}. \label{fig:pairing}}
\end{figure}

\section{Pairing symmetries} 

The strong spin fluctuations in the low-$T$ PM
phase may drive the pairing of electrons.
To determine the symmetry of a possible superconducting state, we
employ a weak-coupling method due to Scalapino {\it et
  al.}~\cite{Scalapino:1986p2491} which has been widely used to study the 
pnictides.~\cite{graser_5b,Graser:2010p2497,Kemper:2010p2492,ql_luo_2010} 

The pairing vertex due to the exchange of spin and charge fluctuations is
obtained within the fluctuation exchange approximation. We have
\beqarray
\hat{\Gamma}^{s}({\bf k},{\bf k}',\omega)
& = &\frac{3}{2}\hat{U}^S\hat{\chi}^{S}({\bf k}-{\bf
    k}',\omega)\hat{U}^{S} +
  \frac{1}{2}\hat{U}^{S}  \notag \\
& &  
  - \frac{1}{2}\hat{U}^C\hat{\chi}^{C}({\bf k}-{\bf k}',\omega)\hat{U}^{C} +
  \frac{1}{2}\hat{U}^{C} \label{eq:pairsing} \\
\hat{\Gamma}^{t}({\bf k},{\bf k}',\omega)
& = & -\frac{1}{2}\hat{U}^S\hat{\chi}^{S}({\bf k}-{\bf
    k}',\omega)\hat{U}^{S} +
  \frac{1}{2}\hat{U}^{S}  \notag \\
& & 
  - \frac{1}{2}\hat{U}^C\hat{\chi}^{C}({\bf k}-{\bf k}',\omega)\hat{U}^{C} +
  \frac{1}{2}\hat{U}^{C} \label{eq:pairtrip}
\eeqarray
for singlet and triplet pairing, respectively. Here $\hat{\chi}^{C}$ is the RPA
irreducible charge susceptibility, which obeys the Dyson equation
$\hat{\chi}^{C} = \hat{\chi}^{(0)} -
\hat{\chi}^{(0)}\hat{U}^{C}\hat{\chi}^{C}$. The non-zero elements of
$\hat{U}^{C}$ are again found in Refs.~\onlinecite{graser_5b,Graser:2010p2497,Kemper:2010p2492}.

Assuming that the dominant scattering occurs close to the
FSs~\cite{graser_5b,Kemper:2010p2492,Graser:2010p2497,ql_luo_2010}, we describe
the scattering of a Cooper pair from the state $({\bf k},-{\bf k})$ on FS $C_{i}$ to the state $({\bf k}',-{\bf k}')$ on FS $C_{j}$
by the projected vertices
\beqarray
\Gamma^{\nu=s,t}_{i,j}({\bf k},{\bf k}') &= &
\sum_{\alpha,\beta,\gamma,\delta}u_{i,\gamma}(-{\bf k})u_{i,\alpha}({\bf
  k})u^{\ast}_{j,\beta}({\bf k}')u^{\ast}_{j,\delta}(-{\bf k}') \notag \\
&& \times \mbox{Re}\left\{\Gamma^{\nu=s,t}_{\alpha,\beta,\gamma,\delta}({\bf k}-{\bf
  k}',\omega=0)\right\} \, .  \label{eq:gammaij}
\eeqarray
The superconducting gap on the FSs is written as 
$\Delta({\bf k}) = \Delta g_{\nu}({\bf k})$, where $g_{\nu}({\bf k})$ is a
dimensionless function describing the momentum dependence, and
$\nu=s$ ($t$) denotes a singlet (triplet) state with even (odd)
parity. $g_{\nu}({\bf k})$ is obtained by solving the eigenvalue problem 
\beq
-\sum_{j}\oint_{C_j}\frac{dk'_\parallel}{4\pi^2v_{F,j}({\bf k}')}
\Gamma^{\nu}_{ij}({\bf k},{\bf k}')g_{\nu}({\bf k}') = \lambda g_{\nu}({\bf k}) \label{eq:geigen}
\eeq
where ${v_{F,i}({\bf k})}$ is the Fermi velocity and
$\lambda$ is a dimensionless coupling strength. The gap function $g_{\nu}({\bf
  k})$ with the largest $\lambda$ has the highest $T_c$, and
hence fixes the symmetry of the pairing state. We use a $192\times192$
  ${\bf k}$-point mesh and $T=50\,$K to
  determine the leading pairing  instability.
Working
  at $T<50\,$K requires a much larger ${\bf k}$-point mesh, especially when
  small momenta are important for the pairing.

Figure~\ref{fig:lambdas} shows the eigenvalues $\lambda$
  of~\eq{eq:geigen} as a function of $U$ for various pairing symmetries  
  at $T=50\,$K. The dominant pairing channel corresponds to a triplet gap
    with $p$-wave symmetry. The gap 
function $g_{x}({\bf k})$ for the $p_x$-wave state is shown in
Fig.~\ref{fig:gap_px}; this state is degenerate with a $p_y$-wave state
with gap function $g_{y}(k_x,k_y) = g_{x}(k_y,k_x)$. In order to maximize the
gap at the FS, these two states can be
combined to form a number of unitary states: (i) ${\bf
  d}({\bf k})= \Delta[g_x({\bf 
    k}) \pm ig_y({\bf k})]\hat{\bf e}_z$, (ii) ${\bf
  d}({\bf k})= \Delta [g_x({\bf k})\hat{\bf e}_x \pm
{g}_y({\bf k})\hat{\bf e}_y]$, and (iii) ${\bf
  d}({\bf k})= \Delta [g_x({\bf
  k})\hat{\bf e}_y \pm {g}_y({\bf k})\hat{\bf e}_x]$. All of these states have
the same gap magnitude, shown in Figure~\ref{fig:gap_comb}, and are hence
degenerate in the present model. The inclusion of additional interactions,
e.g. spin-orbit coupling, may lift this degeneracy and favor one state, as in
Sr$_2$RuO$_4$.~\cite{Sigrist:1991p2454,Mackenzie:2003p2495} The absence of the Knight shift for fields parallel
to the $ab$ plane, and the strong out-of-plane anisotropy, suggests
that the opposite-spin-pairing ${\bf d}({\bf k}) = \Delta[g_x({\bf 
    k}) \pm ig_y({\bf k})]\hat{\bf e}_z$ state is most likely in
  LiFeAs.~\cite{LiFeAs_triplet}

The sub-dominant pairing states are a triplet $f$-wave state and a
singlet $l=6$ state. The latter can become rather strong close to the IC-AF
phase, where  the small-${\bf q}$ IC
fluctuations favor a singlet gap that changes sign over the same small
momentum difference. It is interesting to compare our results with the
related three-orbital model proposed in~\Ref{Daghofer_3orb}, where
scattering across the inner hole pocket also produces a similar ring-shaped
feature in $\chi^{zz}({\bf q})$. The radius of this hole pocket is much
larger than in our model for LiFeAs, however, and there is hence not
enough weight at ${\bf q}=(0,0)$ to support triplet
pairing.~\cite{ql_luo_2010} Instead, the larger radius of the ring feature
favors slowly-varying $s$-wave or $B_{1g}$ singlet gaps. Since triplet SC
is dominant as long as the weight at ${\bf q}=(0,0)$ remains high enough down to the
critical temperature, we conclude that it will be favored by a small inner
hole pocket. 
Other features of the bands play only a secondary role, e.g. by slightly
varying the hopping integrals we can find a much smaller gap on the inner hole
pocket, or an almost constant gap on the electron pockets
apart from very narrow nodal regions.

\section{Conclusions}

In this paper we have presented a coherent picture
  for the properties of LiFeAs driven by magnetic fluctuations. Due to the
poor electron-hole nesting and the shallow hole pockets, we find that 
scattering across the small inner hole pocket dominates the physics, leading
to a ring-shaped feature around $(0,0)$ in the momentum-dependent magnetic
susceptibility. Due to these almost FM processes, triplet $p$-wave
superconductivity is favored over singlet pairing.

\begin{acknowledgments}

We would like to thank G. Martins, A. F. Kemper, S. Graser, I. Mazin, S.-H. Baek,
H.-J. Grafe, F. Hammerath and B. B\"uchner for 
valuable discussions. This research was supported by the DFG under the
priority program 1458 (PMRB, CT, and JvdB) and the Emmy-Noether
program (MD). 
\end{acknowledgments}


\begin{thebibliography}{10}
\bibitem{endn2}
The procedure is similar for the FM phase.
\bibitem{endn3}
Spin fluctuations involving the other pockets are much weaker: a very faint
  ring corresponding to scattering across the outer hole pocket is only visible
  at low temperatures, while there is no clear contribution from the electron
  pockets.
\end{thebibliography}

\end{document}